\documentclass[a4paper]{article}
\usepackage[english]{babel}
\usepackage{amsmath,amsfonts,amssymb}
\usepackage{graphicx}

\begin{document}

\relax
\renewcommand{\theequation}{\arabic{section}.\arabic{equation}}

\def\be{\begin{equation}}
\def\ee{\end{equation}}
\def\bs{\begin{subequations}}
\def\es{\end{subequations}}
\def\calm{{\cal M}}
\def\calk{{\cal K}}
\def\Hc{{\cal H}}
\def\lx{\lambda}
\def\sx{\sigma}
\def\ex{\epsilon}
\def\Lx{\Lambda}

\newcommand{\bett}{\tilde{\beta}}
\newcommand{\alpht}{\tilde{\alpha}}
\newcommand{\dv}{\delta{v}}
\newcommand{\dvvec}{\delta\vec{v}}
\newcommand{\phit}{\tilde{\phi}}
\newcommand{\Phit}{\tilde{\Phi}}
\newcommand{\Gt}{\tilde{G}}
\newcommand{\mt}{\tilde{m}}
\newcommand{\Mt}{\tilde{M}}
\newcommand{\nt}{\tilde{n}}
\newcommand{\Nt}{\tilde{N}}
\newcommand{\Bt}{\tilde{B}}
\newcommand{\Rt}{\tilde{R}}
\newcommand{\rt}{\tilde{r}}
\newcommand{\mut}{\tilde{\mu}}
\newcommand{\mub}{\bar{\mu}}
\newcommand{\vrm}{{\rm v}}
\newcommand{\tl}{\tilde t}
\newcommand{\ttt}{\tilde T}
\newcommand{\rhot}{\tilde \rho}
\newcommand{\ptt}{\tilde p}
\newcommand{\drho}{\delta \rho}
\newcommand{\dpp}{\delta p}
\newcommand{\dphi}{\delta \phi}
\newcommand{\drhot}{\delta {\tilde \rho}}
\newcommand{\dchi}{\delta \chi}
\newcommand{\A}{A}
\newcommand{\B}{B}
\newcommand{\mmu}{\mu}
\newcommand{\mnu}{\nu}
\newcommand{\ii}{i}
\newcommand{\jj}{j}
\newcommand{\jl}{[}
\newcommand{\jr}{]}
\newcommand{\ml}{\sharp}
\newcommand{\mr}{\sharp}

\newcommand{\da}{\dot{a}}
\newcommand{\db}{\dot{b}}
\newcommand{\dn}{\dot{n}}
\newcommand{\dda}{\ddot{a}}
\newcommand{\ddb}{\ddot{b}}
\newcommand{\ddn}{\ddot{n}}
\newcommand{\pa}{a^{\prime}}
\newcommand{\pn}{n^{\prime}}
\newcommand{\ppa}{a^{\prime \prime}}
\newcommand{\ppb}{b^{\prime \prime}}
\newcommand{\ppn}{n^{\prime \prime}}
\newcommand{\fda}{\frac{\da}{a}}
\newcommand{\fdb}{\frac{\db}{b}}
\newcommand{\fdn}{\frac{\dn}{n}}
\newcommand{\fdda}{\frac{\dda}{a}}
\newcommand{\fddb}{\frac{\ddb}{b}}
\newcommand{\fddn}{\frac{\ddn}{n}}
\newcommand{\fpa}{\frac{\pa}{a}}
\newcommand{\fpb}{\frac{\pb}{b}}
\newcommand{\fpn}{\frac{\pn}{n}}
\newcommand{\fppa}{\frac{\ppa}{a}}
\newcommand{\fppb}{\frac{\ppb}{b}}
\newcommand{\fppn}{\frac{\ppn}{n}}
\newcommand{\at}{\tilde{\alpha}}
\newcommand{\pt}{\tilde{p}}
\newcommand{\Ut}{\tilde{U}}
\newcommand{\phidot}{\dot{\phi}}
\newcommand{\rhb}{\bar{\rho}}
\newcommand{\pb}{\bar{p}}
\newcommand{\pbb}{\bar{\rm p}}
\newcommand{\kt}{\tilde{k}}
\newcommand{\kb}{\bar{k}}
\newcommand{\wt}{\tilde{w}}

\newcommand{\dA}{\dot{A_0}}
\newcommand{\dB}{\dot{B_0}}
\newcommand{\fdA}{\frac{\dA}{A_0}}
\newcommand{\fdB}{\frac{\dB}{B_0}}

\def\be{\begin{equation}}
\def\ee{\end{equation}}
\def\bs{\begin{subequations}}
\def\es{\end{subequations}}
\newcommand{\een}{\end{subequations}}
\newcommand{\ben}{\begin{subequations}}
\newcommand{\beq}{\begin{eqalignno}}
\newcommand{\eeq}{\end{eqalignno}}

\def \lta {\mathrel{\vcenter
     {\hbox{$<$}\nointerlineskip\hbox{$\sim$}}}}
\def \gta {\mathrel{\vcenter
     {\hbox{$>$}\nointerlineskip\hbox{$\sim$}}}}

\def\g{\gamma}
\def\mpl{M_{\rm Pl}}
\def\ms{M_{\rm s}}
\def\ls{l_{\rm s}}
\def\l{\lambda}
\def\m{\mu}
\def\n{\nu}
\def\a{\alpha}
\def\b{\beta}
\def\gs{g_{\rm s}}
\def\d{\partial}
\def\co{{\cal O}}
\def\sp{\;\;\;,\;\;\;}
\def\r{\rho}
\def\dr{\dot r}

\def\e{\epsilon}
\newcommand{\NPB}[3]{\emph{ Nucl.~Phys.} \textbf{B#1} (#2) #3}   
\newcommand{\PLB}[3]{\emph{ Phys.~Lett.} \textbf{B#1} (#2) #3}   
\newcommand{\ttbs}{\char'134}        
\newcommand\fverb{\setbox\pippobox=\hbox\bgroup\verb}
\newcommand\fverbdo{\egroup\medskip\noindent%
                        \fbox{\unhbox\pippobox}\ }
\newcommand\fverbit{\egroup\item[\fbox{\unhbox\pippobox}]}
\newbox\pippobox
\def\tr{\tilde\rho}
\def\lb{w}
\def\bbox{\nabla^2}
\def\mt{{\tilde m}}
\def\rct{{\tilde r}_c}

\def \lta {\mathrel{\vcenter
     {\hbox{$<$}\nointerlineskip\hbox{$\sim$}}}}
\def \gta {\mathrel{\vcenter
     {\hbox{$>$}\nointerlineskip\hbox{$\sim$}}}}

\noindent
\begin{flushright}

\end{flushright} 
\vspace{1cm}
\begin{center}
{ \Large \bf Non-linear Matter Spectra 
in Coupled Quintessence \\}
\vspace{0.5cm}
{F. Saracco$^{(1)}$, M. Pietroni$^{(2)}$, 
N. Tetradis$^{(3)}, $V. Pettorino$^{(4,5)}$, G. Robbers$^{(6)}$ } 
\end{center}
\vspace{0.6cm}
(1) {\it INFN, Sezione di Firenze, via G. Sansone 1, Sesto Fiorentino, Firenze 50019, Italy} 
\\
(2) {\it INFN, Sezione di Padova, via Marzolo 8, Padova 35131, Italy}
\\
(3) {\it Department of Physics, University of Athens, 
University Campus, Zographou 157 84, Greece
\\
(4) {\it Institut f\"ur Theoretische Physik, Universit\"at Heidelberg, Philosophenweg 16,
Heidelberg 69120, Germany}
\\
(5) Italian Academy for Advanced Studies in America,
Columbia University,
1161 Amsterdam Avenue,
New York, NY 10027, USA
\\
(6) {\it Max-Planck-Institut f\"ur Astrophysik, Karl-Schwarzschild-Strasse 1, Garching bei M\"unchen 85748, Germany}
} 
\vspace{1cm}
\abstract{
We consider cosmologies in which a dark-energy scalar field interacts with cold dark matter. The growth of perturbations is followed beyond the linear level by means of the time-renormalization-group method, which is extended to describe a multi-component matter sector. Even in the absence of the extra interaction, a scale-dependent bias is generated as a consequence of the different initial conditions for baryons and dark matter after decoupling. The effect is  enhanced significantly by the extra coupling and can be at the 2-3 percent level in the range of scales of baryonic acoustic oscillations. We compare our results with N-body simulations, finding very good agreement.
}

\newpage

\section{Introduction}
\setcounter{equation}{0}

The  power spectrum of matter perturbations reflects the evolution of the Universe since
the time of matter-radiation equality. For given initial conditions, determined by the 
primordial spectrum (usually assumed to be scale invariant), the growth of perturbations depends on the cosmological scenario. 
The calculation of the present matter power spectrum can constrain this scenario through the 
comparison of the deduced spectrum with the observed large-scale structure. A major technical
difficulty in the realization of such a program is the failure of linear perturbation theory to describe
present-day fluctuations with characteristic length scales below roughly 100 Mpc.  
At length scales below about 10 Mpc, the evolution is highly
non-linear, so that only numerical N-body simulations can capture the dynamics of the formation 
of galaxies and clusters of galaxies. Fluctuations with length scales of 
around 100 Mpc fall within the mildly non-linear regime, for which analytical methods
have been developed. 
These scales (corresponding to wavenumbers in the $0.03-0.25\,h$/Mpc range)
are of particular interest, because they correspond to the sound horizon at
decoupling, which can be determined by reconstructing the  oscillatory behavior of the
matter power spectrum due to baryonic acoustic oscillations (BAO).

The various analytical methods \cite{CrSc1}--\cite{matsubara} essentially amount to 
resummations of subsets of perturbative diagrams of arbitrarily high order, in a way analogous to the 
renormalization group (RG). 
In this work we shall follow the approach of  \cite{Max2}, named time-RG or TRG. In the context of the RG
the various observables depend on 
a characteristic energy scale, and evolve as this scale is varied. The TRG 
uses time as the flow parameter that describes the evolution of physical quantities,
such as the spectrum of perturbations. The method is characterized by conceptual
simplicity. It has been applied to ${\rm \Lambda CDM}$ and quintessence cosmologies \cite{Max2}, as well as
models with massive neutrinos \cite{lesgourgues}. 

The fundamental equations in the TRG approach are the ``equations of motion'', i.e. the
continuity, Euler and Poisson equations. From these, equations can be derived for the time evolution of
correlation functions for the density and velocity fields.  The various spectra are obtained through appropriate
Fourier transforms of the correlation functions. The method results in a coupled infinite system of integro-differential
equations for the time evolution of the spectrum, bispectrum etc. 
The crucial approximation, that makes a solution possible, is to neglect the effect of the higher-order correlation functions
in the evolution equations of the lower-order ones. The calculations performed so far take into account the spectrum and
bispectrum and set the higher-level spectra to zero. 

The procedure of truncating the system of equations is commonly employed 
in the applications of the Wilsonian RG to field theory or statistical physics. (For a review, see \cite{erg}.)
The accuracy of the calculation can be determined either by enlarging the truncated system (by including the trispectrum,
for example) and examining the stability of the results, or by comparing with alternative methods. The second 
approach is often followed, because enlarging the truncation can increase the complexity of the calculation considerably.
In the case of the TRG, the agreement with results from N-body simulations for ${\rm \Lambda CDM}$ has been
confirmed \cite{Max2}. Also, a comparative analysis of several analytical methods, using N-body simulations as 
a reference, has been carried out in ref. \cite{carlson}. The study  
demonstrates that the TRG remains accurate at the 1-2\% level over the whole BAO range  at all redshifts. 

In fig. 5 of ref. \cite{carlson} the deviation of the TRG prediction from a reference spectrum 
derived through simulations  for ${\rm \Lambda CDM}$ is depicted. At a redshift $z=1$ the deviation is at the
1\% level, or smaller, over the whole depicted range $0\leq k \leq 0.15\,h$/Mpc. For $z=0$ the deviation may exceed
1\% for some values of $k$, but stays below 2\% over the whole depicted range. Based on these findings
and the analysis of \cite{Max2} we estimate an accuracy of 1\% for $z=1$ and 2\% for $z=0$ over the range
$0\leq k \leq 0.2\,h$/Mpc. The accuracy of 1-loop standard perturbation theory (SPT) can be inferred from fig. 1 of ref. \cite{carlson}.
At $z=1$ the accuracy is at the 1\% level for $0\leq k \leq 0.1\,h$/Mpc, while at $z=0$ it is at the 2\% level for 
$0\leq k \leq 0.05\,h$/Mpc. We depict these ranges in figs. \ref{ddcc}--\ref{bias1} of the current paper. 
The additional approximations that we make in this paper for the study of models with non-zero coupling between 
dark matter and dark energy induce uncertainties at the sub-percent level. We discuss this issue in detail
in subsection 3.1. For this reason the level of accuracy of our results for the power spectra in the coupled case
is expected to be similar to that for ${\rm \Lambda CDM}$. Its magnitude is set by the truncations in the 
evolution equations (the omission of the effect of the trispectrum and higher spectra), which are similar  
in all models.

The purpose of the present work is threefold:
\\
1) We extend the formalism to more complicated models. We introduce two modificiations to 
previous studies: a) Within the matter sector we allow for an arbitrary number of species with independent
spectra. These include baryonic matter (BM) and cold (pressureless) dark matter (CDM). One may also consider 
contributions from massive neutrinos etc. b) We allow for an interaction between CDM and dark energy (DE). 
We consider a class of
quintessence models, in which there is direct coupling between CDM and the quintessence field. The
form of the interaction is a generalization of the universal coupling to all species present in scalar-tensor theories in the Einstein frame. 
It is modelled through the dependence of the mass of the CDM particles on the quintessence 
field \cite{wetcosmon}-\cite{baccigalupi}. 
\\
2) We test the accuracy of the method in this enlarged framework by comparing with available N-body simulations. 
We perform our numerical analysis for a model for which results from simulations are given in ref. \cite{baldi}. 
In the context of coupled quintessence, the cosmological evolution can be very diverse
\cite{amendola,baccigalupi,perturbations1,perturbations2}. 
It is very time-consuming to study exhaustively every model through N-body simulations.
Our approach provides an alternative method, which can be much faster, while retaining the necessary accuracy.
It is also important to note that, while the N-body simulations 
are highly accurate at the length scales of galaxies and clusters of galaxies, they are less accurate in the BAO range
because of the required large volumes. On the other hand, analytical methods, like ours, are more accurate in
the quasi-linear regime of large length scales. The two  approaches can be viewed as complementary. 
\\
3) We provide predictions for observables for which non-linear corrections can be important. As such, we study the
bias between dark and baryonic matter in the BAO range for models of coupled quintessence.

The couplings between the matter sector and DE are constrained by observations. For the BM-DE coupling, the bound from 
the Cassini spacecraft \cite{Cassini} limits its order of magnitude to be below $10^{-3}$. As such small 
couplings produce negligible effects on the power spectrum, we assume that the BM-DE coupling is exactly zero.  
The coupling between CDM and DE is constrained by various considerations, such as the modification of the spectrum
of the cosmic microwave background (CMB) or that of the matter distribution. A common feature of the class of models 
we are considering is that the presence 
of an additional long-range force between CDM particles, induced by the DE field, modifies their clustering properties.
The various observable consequences have been discussed in the literature \cite{largescale}--\cite{baldi}. 
The strength of the CDM-DE interaction is constrained through the comparison with the observed CMB and matter 
spectra. It has been shown that, for particular models,
the CDM-DE coupling must be considerably smaller than the gravitational one \cite{bean,lavacca}. 
Such constraints cannot be considered generic, because the evolution of
the cosmological background and the perturbations around it varies considerably from model to model.
We work within the model of \cite{baldi}, because our main objective is to compare with the results of 
N-body simulations presented there. The couplings that we consider are roughly consistent with the bounds 
deduced in \cite{bean,lavacca} for a model similar to ours. 

In this work we follow a novel approach for the derivation of the fundamental equations.
We derive the continuity, Euler and Poisson equations on an expanding background, starting from
the conservation of the energy-momentum tensor. 
In order to cast these equations in a form that generalizes the standard expressions on a static background,
we need to impose a certain hierarchy between the density perturbations, the velocity field and the potentials.
Our derivation makes it straightforward to generalize these equations in future studies 
in order to take into account non-zero pressure and
higher-order terms. 
Next, we derive the system of differential equations for the spectrum and bispectrum, within
a truncation that neglects higher-level spectra. We integrate these equations numerically in order to
produce the non-linear spectra at low redshift and compare with  the results of N-body simulations. 
We study in detail the difference, usually characterized as 
bias, between the spectra of dark and baryonic matter. We find that the CDM-DE coupling enhances significantly the bias
of the decoupled case.

In the following section we derive the evolution equations for the spectra of dark and baryonic matter.
The details of the derivation for the case of one massive component are given in appendix A. 
The generalization for an arbitrary number of massive components is presented in appendix B. 
The results of the numerical integration of the evolution equations are presented in section 3. We compare them with the results of
N-body simulation. We also discuss in detail the form of the bias in the BAO range.

\section{Dark matter coupled to dark energy}
\setcounter{equation}{0}

\subsection{Non-linear evolution equations for the perturbations}

We assume that the energy density of the Universe 
receives significant contributions from three components: 
a) standard baryonic matter (BM); 
b) a species of weakly interacting, massive particles, which we identify with
cold dark matter (CDM); and  
c) a slowly varying, classical scalar field $\phi$, whose contribution to the energy
density is 
characterized as dark energy (DE).
We also consider the possibility that there is a direct coupling 
between the CDM particles and the 
scalar field. 
Its equation of motion takes the form
\be
\frac{1}{\sqrt{-g}}\frac{\partial}{\partial x^\mu}
\left(\sqrt{-g}\,\,g^{\mu\nu}\frac{\partial \phi}{\partial x^\nu}
\right)
=-\frac{dU}{d\phi}+{\beta(\phi)}\,\, \left( T_{CDM} \right)^\mu_{~\mu}.
\label{eomphia} \ee
We normalize all dimensionful quantities, such as the
scalar field, with respect to the reduced Planck mass $M=(8\pi G)^{-1/2}$. 
The full $M$-dependence is displayed explicitly
in appendix A. Our normalization here is equivalent to setting $M=1$. 
Equation (\ref{eomphia}) can be obtained if we assume that the 
mass $m$ of the particles has a dependence on $\phi$ \cite{brouzakis}. 
Then we have $\beta(\phi)=-{d\ln m(\phi)}/{d\phi}$. 
In order to be consistent with the stringent observational constraints for the baryonic sector, we assume that
the interaction with the DE scalar field is confined to the CDM sector. The BM has no
direct coupling to $\phi$. 

For the metric, we consider an ansatz of the form
\be
ds^2=a^2(\tau)\left[
\left(1+2\Phi(\tau,\vec{x}) \right)d\tau^2
-\left(1-2\Phi(\tau,\vec{x}) \right) d\vec{x}\, d\vec{x} \right].
\label{metric} \ee
We assume that the Newtonian potential $\Phi$ is weak, $\Phi \ll 1$, and  
that the field $\phi$ can be decomposed as
\be
\phi(\tau,\vec{x})=\bar{\phi}(\tau)+\dphi(\tau,\vec{x}),
\label{decomphi} \ee
with $\dphi/\bar{\phi} \ll 1$. In general,
$\bar{\phi}={\cal O}(1)$ in units of $M$.
The magnitude of the fluctuations of $\phi$ is 
expected to be similar to that of the gravitational field $\Phi$.
The reason is that the source for both is the dark matter density, to 
which they couple with comparable strength (as will be apparent in the following). 
Finally, the density can be decomposed as
\be
\rho(\tau,\vec{x})=\bar{\rho}(\tau)+\drho(\tau,\vec{x}).
\label{decomrho} \ee
We allow for significant
density fluctuations, even though our analysis is not applicable when they
are much larger than the background density.  
Our aim is to take into account the effect of the local velocity field
$\dvvec$,
when this becomes significant because of large field gradients. For 
subhorizon perturbations with momenta $k\gg \Hc=\dot{a}/a$, 
the linear analysis predicts 
$|\dvvec| \sim (k/\Hc) \Phi \sim (\Hc/k)(\drho/\bar{\rho})$. 
A consistent expansion scheme can be obtained if we assume that 
$\Phi \ll |\dvvec | \ll 1$. 
Including the density perturbations, our assumptions can be summarized in the  
hierarchy of
scales: $\Phi, \dphi/\bar{\phi} \ll |\dvvec | \ll \drho/\bar{\rho}  \lesssim 1$. 
At the linear level, 
we have $\dvvec^2 \sim \Phi (\drho/\bar{\rho})$. We assume that such a relation
holds at the non-linear level as well, within the range of applicability of 
our scheme.
The velocity field is driven by the spatial
derivatives of the potentials $\Phi, \dphi/\bar{\phi}$. As we are dealing
with subhorizon perturbations, it is consistent to make
the additional assumption that the spatial derivatives of $\Phi, \dphi$
dominate over their time derivatives. 
The predictions of the linear analysis allow us to make a more quantitative
statement. We assume that a spatial derivative acting on $\Phi$, $\dphi$ or 
$\dvvec$ increases the position of that quantity in the hierarchy by one 
level. In this sense $\vec{\nabla} \Phi$ is comparable to $\dvvec$, while
$\nabla^2 \Phi$ is comparable to $\bar{\rho}$. 

With the above assumptions, one can derive the equations 
that describe the evolution of the Universe. For one non-relativistic 
species, the derivation is presented in
appendix A. It is generalized to two species in appendix B. 
The evolution of the homogeneous background is described by 
\begin{subequations}
\begin{align}
\mathcal{H}^{2}=&\dfrac{1}{3}\big[a^{2}\sum_{i=1,2}\bar{\rho}_{i}
+\frac{1}{2}{\dot{\bar{\phi}}^{2}}+a^{2} U(\bar{\phi})\big]
\equiv\dfrac{1}{3}{a^{2}\rho_{tot}}\\
\dot{\bar{\rho}}_{i}+3\mathcal{H}\bar{\rho}_{i}
=&-\beta_{i}\,\dot{\bar{\phi}}\bar{\rho}_{i}\\
\ddot{\bar{\phi}}+2\mathcal{H}\dot{\bar{\phi}}=&-{a^{2}}
\left(\dfrac{dU}{d\phi}(\bar{\phi})-\sum_{i=1,2}\beta_{i}\,\bar{\rho}_{i}\right),
\label{25c} \end{align}
\end{subequations}
where we have defined $\rho_{tot}\equiv\sum_{i}\bar{\rho}_{i}
+{\dot{\bar{\phi}}^{2}}/{(2a^{2})}+U(\bar{\phi})$. For the CDM we have a non-zero constant
$\beta_1=\beta$, while for BM, because of the strong constraints from solar system tests of General Relativity \cite{Cassini}, 
we set $\beta_2=0$.
 
We describe the perturbations in terms of the scalar-field perturbation $\delta\phi$,
the Newtonian potential $\Phi$, the density perturbations $\delta\rho_i$ and the 
velocity fields $v_i$.
We have two Poisson equations 
\begin{subequations}
 \begin{align}
  \nabla^{2}\delta\phi=&-{a^{2}\sum_{i}\beta_{i}\delta\rho_{i}}
\equiv -3\sum_{i}\beta_{i}\mathcal{H}^{2}\Omega_{i}\delta_{i} \label{poiss1} \\
\nabla^{2}\Phi=&\frac{1}{2}a^{2}{\sum_{i}\delta\rho_{i}}
\equiv\dfrac{3}{2}\mathcal{H}^{2}\sum_{i}\Omega_{i}\delta_{i},
\label{poiss2} \end{align}
\end{subequations}
with 
$\Omega_{i}(\tau)
\equiv{\bar{\rho}_{i}\,a^{2}}/{(3\mathcal{H}^{2})}$,
and the continuity and Euler equations
\begin{subequations}
 \begin{align}
{\delta\dot{\rho}}_{i}+3\mathcal{H}\delta\rho_{i}+\vec{\nabla}
[(\bar{\rho}_{i}+\delta\rho_{i})\delta{\vec{v}_i}]=&-\beta_{i}\dot{\bar{\phi}}
\delta\rho_{i}\label{a}\\
 \delta \dot{\vec{v}}_{i}+(\mathcal{H}-\beta_{i}\dot{\bar{\phi}})
\delta \vec{v}_{i}+(\delta{\vec{v}}_{i}\cdot\vec{\nabla})\delta{\vec{v}}_{i}
=&-\vec{\nabla}\Phi+\beta_{i}\vec{\nabla}\delta\phi\label{b}. 
 \end{align}
\end{subequations}

\subsection{The CDM-BM quadruplet and the power spectra}\label{model}
The evolution equations are expressed in their most 
useful form in terms of the density contrasts 
$\delta_{i}\equiv{\delta\rho_{i}}/{\bar{\rho}_{i}}\lesssim 1$
and 
$\theta_{i}(\textbf{k}, \tau)\equiv\vec{\nabla}\cdot\vec{\delta v_{i}}(\textbf{k}, \tau)$.
For the Fourier transformed quantities, 
we obtain from eq. (\ref{a})
\begin{equation}\label{delta}
\dot{\delta_{i}}(\textbf{k}, \tau)+\theta_{i}(\textbf{k}, \tau)
+\int d^{3}\textbf{k}_{1} \,d^{3}\textbf{k}_{2}\,
\delta_{D}(\textbf{k}-\textbf{k}_{1}-\textbf{k}_{2})\,
\alpht(\textbf{k}_{1}, \textbf{k}_{2})\,
\theta_{i}(\textbf{k}_{1}, \tau)\,
\delta_{i}(\textbf{k}_{2}, \tau)
=0,
\end{equation}
where 
\begin{equation}
\alpht(\textbf{k}_{1}, \textbf{k}_{2})=\dfrac{\textbf{k}_{1}\cdot(\textbf{k}_{1}
+\textbf{k}_{2})}{k_{1}^{2}}. 
\end{equation}
Eqs. (\ref{b}), (\ref{poiss1}), (\ref{poiss2}) give
\begin{equation}\label{theta}
\begin{split}
\dot{\theta}_{i}(\textbf{k}, \tau)+
&(\mathcal{H}-\beta_{i}\dot{\bar{\phi}})\theta_{i}(\textbf{k}, \tau)
+\dfrac{3\mathcal{H}^{2}\sum_{j}\Omega_{j}(2\beta_i\beta_{j}+1)\delta_{j}
(\textbf{k}, \tau)}{2}\\
+&\int d^{3}\textbf{k}_{1}\, d^{3}\textbf{k}_{2}\,
\delta_{D}(\textbf{k}-\textbf{k}_{1}-\textbf{k}_{2})\,
{\bett}(\textbf{k}_{1}, \textbf{k}_{2})\,
\theta_{i}(\textbf{k}_{1}, \tau)\,
\theta_{i}(\textbf{k}_{2}, \tau)=0,
\end{split}
\end{equation}
where 
\begin{equation}
 \bett(\textbf{k}_{1}, \textbf{k}_{2})=\dfrac{(\textbf{k}_{1}+\textbf{k}_{2})^{2}
\textbf{k}_{1}\cdot\textbf{k}_{2}}{2 k_{1}^{2}k_{2}^{2}}.
\end{equation}

In appendix B we discuss the above equations for an arbitrary number of non-relativistic
species. Here we concentrate on the case of interest, i.e. CDM coupled to $\phi$ and BM
only gravitationally coupled.
We define the quadruplet
\begin{equation}
 \left(
\begin{array}{c}
\varphi_{1}(\textbf{k}, \eta)\\ \\ \varphi_{2}(\textbf{k}, \eta)\\ \\ 
\varphi_{3}(\textbf{k}, \eta)\\ \\ \varphi_{4}(\textbf{k}, \eta)\end{array}
\right)
=e^{-\eta}\left(
\begin{array}{c}
\delta_{CDM}(\textbf{k}, \eta)\\ \\-\dfrac{\theta_{CDM}(\textbf{k}, \eta)}{\mathcal{H}}\\ \\ 
\delta_{BM}(\textbf{k}, \eta)\\ \\-\dfrac{\theta_{BM}(\textbf{k}, \eta)}{\mathcal{H}}
\end{array}
\right),
\label{quadruplet}
\end{equation}
where $\eta=\ln a(\tau)$.  
This allows us to bring 
eqs. (\ref{delta}), (\ref{theta}) 
in the form \cite{CrSc1,Max1,Max2} 
\begin{equation}\label{arghh}
\partial_{\eta}\varphi_{a}(\textbf{k}, \eta)+\Omega_{ab}\varphi_{b}(\textbf{k}, \eta)
=e^{\eta}\gamma_{abc}(\textbf{k}, -\textbf{k}_{1}, -\textbf{k}_{2})
\varphi_{b}(\textbf{k}_{1}, \eta)\varphi_{c}(\textbf{k}_{2}, \eta).
\end{equation}
The indices $a,b,c$ take values $1,\ldots, 4$.
The values 1,2
characterize CDM density and velocity perturbations, while 3,4 refer
to BM quantities.  
Repeated momenta are integrated over, while repeated indices are summed over. 
The functions $\gamma$, that determine effective vertices, 
are analogous to those employed in \cite{Max1,Max2}.
The non-zero components are
\begin{equation}
\begin{split}
\gamma_{121}(\textbf{k},\textbf{k}_{1}, \textbf{k}_{2})&
=\dfrac{\alpht(\textbf{k}_{1}, \textbf{k}_{2})}{2}
\delta_{D}(\textbf{k}+\textbf{k}_{1}+\textbf{k}_{2})
=\gamma_{112}(\textbf{k}, \textbf{k}_{2}, \textbf{k}_{1})\\
\gamma_{222}(\textbf{k},\textbf{k}_{1}, \textbf{k}_{2})
&=\bett(\textbf{k}_{1}, \textbf{k}_{2})\ 
\delta_{D}(\textbf{k}+\textbf{k}_{1}+\textbf{k}_{2})\\
\gamma_{343}(\textbf{k},\textbf{k}_{3}, \textbf{k}_{4})
&=\dfrac{\alpht(\textbf{k}_{3}, \textbf{k}_{4})}{2}
\delta_{D}(\textbf{k}+\textbf{k}_{3}+\textbf{k}_{4})
=\gamma_{334}(\textbf{k}, \textbf{k}_{4}, \textbf{k}_{3})\\
\gamma_{444}(\textbf{k},\textbf{k}_{3}, \textbf{k}_{4})
&=\bett(\textbf{k}_{3}, \textbf{k}_{4})\ 
\delta_{D}(\textbf{k}+\textbf{k}_{3}+\textbf{k}_{4}).
\end{split}
\label{vertex}
\end{equation}
The $\Omega$-matrix is
\be
 \Omega(\eta)=\left(
\begin{array}{cccc}
 1 & -1 & 0 & 0
\\ \\
-\dfrac{3}{2}\Omega_{CDM}(2\beta^{2}+1) & 2-\beta\bar{\phi}'+\dfrac{\mathcal{H}'}{\mathcal{H}} & -\dfrac{3}{2}\Omega_{BM} & 0
\\ \\ 
0 & 0 & 1 & -1
\\ \\
-\dfrac{3}{2}\Omega_{CDM} & 0 & -\dfrac{3}{2}\Omega_{BM} & 2+\dfrac{\mathcal{H}'}{\mathcal{H}}
\end{array}
\right),
\label{omegafour}
\ee
where a prime denotes a derivative with respect to $\eta$.

The next step is to derive evolution equations for the power spectra.  
The spectra, bispectra and trispectra are defined as
\begin{equation}
\begin{split}
 \langle\varphi_{a}(\textbf{k}, \eta)\varphi_{b}(\textbf{q}, \eta)\rangle
\equiv&
\delta_{D}(\textbf{k}+\textbf{q}) P_{ab}(\textbf{k}, \eta)\\
\langle\varphi_{a}(\textbf{k}, \eta)\varphi_{b}(\textbf{q}, \eta)\varphi_{c}(\textbf{p}, \eta)\rangle
\equiv&
\delta_{D}(\textbf{k}+\textbf{q}+\textbf{p}) B_{abc}(\textbf{k}, \textbf{q},\textbf{p},\eta)\\
\langle\varphi_{a}(\textbf{k}, \eta)\varphi_{b}(\textbf{q}, \eta)\varphi_{c}(\textbf{p}, \eta)\varphi_{d}(\textbf{r}, \eta)\rangle
\equiv&
\delta_{D}(\textbf{k}+\textbf{q})\delta_{D}(\textbf{p}+\textbf{r}) P_{ab}(\textbf{k}, \eta) P_{cd}(\textbf{p}, \eta)\\
&+\delta_{D}(\textbf{k}+\textbf{p})\delta_{D}(\textbf{q}+\textbf{r}) P_{ac}(\textbf{k}, \eta) P_{bd}(\textbf{q}, \eta)\\
&+\delta_{D}(\textbf{k}+\textbf{r})\delta_{D}(\textbf{q}+\textbf{p}) P_{ad}(\textbf{k}, \eta) P_{bc}(\textbf{q}, \eta)\\
&+\delta_{D}(\textbf{k}+\textbf{p}+\textbf{q}+\textbf{r})Q_{abcd}(\textbf{k},\textbf{p},\textbf{q},\textbf{r},\eta).
\end{split}
\label{spectra} \end{equation}
In appendix B we summarize the derivation of the evolution equations for an arbitrary number of species. 
The essential approximation that we have to make in order to obtain a closed 
system is to neglect the effect of the trispectrum on the
evolution of the bispectrum. In this way we obtain 
\begin{eqnarray}
\partial_{\eta}P_{ab}(\textbf{k}, \eta)&=&-\Omega_{ac}P_{cb}(\textbf{k}, 
\eta)-\Omega_{bc}P_{ac}(\textbf{k}, \eta)
\nonumber \\
&&+e^{\eta}\int d^{3}q \big[\gamma_{acd}(\textbf{k},-\textbf{q}, \textbf{q}-\textbf{k})
B_{bcd}(\textbf{k},-\textbf{q}, \textbf{q}-\textbf{k})
\nonumber \\
&&+\gamma_{bcd}(\textbf{k},
-\textbf{q}, \textbf{q}-\textbf{k})B_{acd}(\textbf{k},-\textbf{q}, \textbf{q}
-\textbf{k})\big],
\label{spectev1}\\
 \partial_{\eta}B_{abc}(\textbf{k},-\textbf{q}, \textbf{q}-\textbf{k})&=&
-\Omega_{ad}B_{dbc}(\textbf{k},-\textbf{q}, \textbf{q}-\textbf{k})-\Omega_{bd}B_{adc}
(\textbf{k},-\textbf{q}, \textbf{q}-\textbf{k})-\Omega_{cd}B_{abd}(\textbf{k},
-\textbf{q}, \textbf{q}-\textbf{k})
\nonumber \\
&&+2e^{\eta}\big[\gamma_{ade}(\textbf{k},-\textbf{q}, \textbf{q}-\textbf{k})
P_{db}(\textbf{q}, \eta)P_{ec}(\textbf{k}-\textbf{q}, \eta)
\nonumber \\
&&+\gamma_{bde}(-\textbf{q},\textbf{q}-\textbf{k}, \textbf{k})P_{dc}(\textbf{k}
-\textbf{q}, \eta)P_{ea}(\textbf{k}, \eta)
\nonumber \\
&&+\gamma_{cde}(\textbf{q}-\textbf{k}, 
\textbf{k}, -\textbf{q})P_{da}(\textbf{k}, \eta)P_{eb}(\textbf{q}, \eta)\big].
\label{spectev2}
\end{eqnarray}

The formal solution of the above system is given by
\begin{eqnarray}
 P_{ab}({\bf k}\,,\eta) &=& g_{ac}({\bf k}\,,\eta,0)  \, g_{bd}({\bf k}\,,\eta,0)  P_{cd}({\bf k}\,,\eta=0) \nonumber\\
&&  +\int_0^\eta d\eta^\prime e^{\eta^\prime} \int d^3 q \,g_{ae}({\bf k}\,,\eta,\eta^\prime) g_{bf}({\bf k}\,,\eta,\eta^\prime) \nonumber\\
&& \quad\times\left[  \gamma_{ecd}({\bf k},\,{\bf -q},\,{\bf q-k})\,B_{fcd}({\bf k},\,{\bf -q},\,{\bf q-k};\,\eta^\prime)\right.
\nonumber\\
&& \quad \quad
+ \left.\gamma_{fcd}({\bf k},\,{\bf -q},\,{\bf q-k})\,B_{ecd}({\bf k},\,{\bf -q},\,{\bf q-k};\,\eta^\prime)\right]\,,
\label{formalsol1} \\
B_{abc}({\bf k},\,{\bf -q},\,{\bf q-k};\,\eta)&=&
g_{ad}({\bf k}\,,\eta,0)g_{be}({\bf -q}\,,\eta,0) g_{cf}({\bf q-k}\,,\eta,0)B_{def}({\bf k},\,{\bf -q},\,{\bf q-k};\,\eta=0)
\nonumber\\
&&+2 \int_0^\eta d\eta^\prime e^{\eta^\prime} \,g_{ad}({\bf k}\,,\eta,\eta^\prime) g_{be}({\bf -q}\,,\eta,\eta^\prime) g_{cf}({\bf q-k}\,,\eta,\eta^\prime)\nonumber\\
&&\quad \times\left[ \gamma_{dgh}({\bf k},\,{\bf -q},\,{\bf q-k})P_{eg}({\bf q}\,,\eta^\prime)P_{fh}({\bf q-k}\,,\eta^\prime)\right.\nonumber\\
&& \quad \quad+ \gamma_{egh}({\bf -q},\,{\bf q-k},\,{\bf k})P_{fg}({\bf q-k}\,,\eta^\prime)P_{dh}({\bf k}\,,\eta^\prime)\nonumber\\
&& \quad \quad\left.
+ \gamma_{fgh}({\bf q-k},\,{\bf k},\,{\bf -q})P_{dg}({\bf k}\,,\eta^\prime)P_{eh}({\bf q}\,,\eta^\prime)
\right]\,,
\label{formalsol2}
\end{eqnarray}
where $g_{ab}({\bf k}\,,\eta,\eta^\prime)$ is the linear propagator, which gives the evolution of the field at the linear level: $\varphi^L_a({\bf k}, \eta) = g_{ab}({\bf k}\,,\eta,\eta^\prime) \varphi^L_b({\bf k}, \eta^\prime)$.

The solutions can be expanded in powers of the interaction vertex $\gamma_{abc}$, in order to 
establish the connection with perturbation theory \cite{Max2}.
The lowest order, corresponding to linear theory, is obtained by setting $\gamma_{abc}=0$. The linear spectrum $ P^L_{ab}$ 
and bispectrum $ B^L_{abc}$ 
are given by the first line of each of the above equations. 
The $O(\gamma)$ correction for the bispectrum is obtained by inserting $ P^L_{ab}$ in place of $P_{ab}$ in the r.h.s. of eq.~(\ref{formalsol2}). Inserting the bispectrum at this order in eq.~(\ref{formalsol1}) generates the $O(\gamma)$ and $O(\gamma^2)$ contributions to the power spectrum. 
At this order, the result for the power spectrum reproduces exactly 
the  one-loop expression in standard perturbation theory (SPT) \cite{bernardeau}.
Iterating the procedure generates the higher-order corrections. However, differences with perturbation theory
arise at higher orders, because of  the approximation $Q_{abcd}=0$ that we have made in deriving the evolution equations 
for the power spectrum and bispectrum.

\section{Numerical analysis}
\setcounter{equation}{0}

\subsection{Approximations}

The presence of two massive species (BM and CDM) complicates the structure of the equations compared to the case where they are 
treated as a single fluid,  discussed in \cite{Max2}. The full system of eqs.~(\ref{spectev1}), (\ref{spectev2}) contains 74 equations, namely, 10 for the power spectra and 64 for the bispectra, compared to the 11 equations of the single-matter case. An accurate 
calculation also requires the discretization of the $k$-space with at least 500 points. Taking into account that the bispectra depend 
on three  external momenta, it is apparent that the necessary computing power is significant. 

The system can be reduced if additional approximations are made, based on the following observations:
\begin{itemize}
\item
The dynamical vertices of 
eq.~(\ref{vertex}) do not mix the CDM components with the BM ones. The coupling between the two type of components 
is entirely due to the linear part of the equations, and especially to the $\Omega_{23}$ and $\Omega_{41}$ entries of eq.~(\ref{omegafour}), through which the 1,2 and 3,4 indices are mixed. These originate in the Poisson equation, 
in which the fluctuations of all the matter species contribute universally to the gravitational potential. 
\item
The ratio of BM and CDM density perturbations is usually characterized as bias: $b =\delta_{BM}/\delta_{CDM}$. 
If at early times $b$ is independent of ${k}$, the subsequent linear evolution preserves this independence, so that 
$b$ is only a function of $\eta$. 
At the linear level, the density-velocity and velocity-velocity spectra are proportional to the density-density ones, with 
$k$-independent proportionality factors. These factors are appropriate powers of the linear growth functions, so that
the growing modes are selected \cite{Max2}. 
\item
The effect of the CDM-DE coupling $\beta$ on the bias can be estimated analytically in the linear approximation.
For $\beta=0$
the evolution equations for CDM and BM are the same, so that, for identical initial conditions, we have
$P_{11}=P_{33}=P_{13}$. For $\beta\not= 0$, the growing mode solution of the linearized equations has the 
form $\varphi_a(\textbf{k}, \eta)=\varphi(\textbf{k}, \eta)\left[ 1, f(\eta), b(\eta),b(\eta)f(\eta)\right]$, where
$f(\eta)$, $b(\eta)$ are solutions of the system
\begin{eqnarray}
\dfrac{3}{2}\Omega_{CDM}(2\beta^{2}+1) +\dfrac{3}{2}\Omega_{BM} b - \frac{3}{2}\frac{1}{b}\Omega_{CDM}-
\frac{3}{2}\Omega_{BM} +\beta\bar{\phi}' f&=&0
\label{grow1} \\
f'+\dfrac{\mathcal{H}'}{\mathcal{H}}f+f+f^2
-\dfrac{3}{2}\frac{1}{b}\Omega_{CDM} - \frac{3}{2}\Omega_{BM}&=&0,
\label{grow2}
\end{eqnarray}
with the prime denoting derivatives with respect to $\eta=\ln a(\tau)$. For $\beta=0$, we have $b=b_0=1$ and
$f=f_0(\eta)$, with $f_0(\eta)$ the growth function of the corresponding decoupled model.  
The corrections for $\beta\not= 0$ are ${\cal O}(\beta^2)$. This is obvious for those arising from the  
first term in eq. (\ref{grow1}). The last term in the same equation has a similar effect because the 
evolution of the field is given by eq. (\ref{25c}). In all models in which the CDM-DE coupling affects the cosmological
evolution, the two terms in the r.h.s. of eq. (\ref{25c}) are comparable, so that 
$\bar{\phi}'={\cal O}(\beta)$.

\item 
The non-linear corrections induce a ${k}$-dependence at low redshifts and large $k$. Again this is an  ${\cal O}(\beta^2)$ effect.
This can be verified in the context of the loop 
expansion, as the vertices are $\beta$-independent, while the propagators for CDM and BM differ by terms of ${\cal O}(\beta^2)$.

\item
Finally, there is a bias induced by the initial conditions for the CDM and BM spectra, which are not identical at the
end of the decoupling era. The magnitude of this effect can be deduced from our subsequent analysis, and is apparent in  
fig. \ref{bias1}, in which the bias in the BAO region is depicted  at a redshift $z=1.12$. The bias
deviates from 1 at a level smaller than 2\% for the decoupled scenario ($\beta=0$).
We have checked that at $z=0$ the effect is below 1\%.  

\item
The estimates on the $\beta$-dependence of the bias are also confirmed by fig. \ref{bias1}: Within the BAO
range, the bias factor $b$ receives corrections of ${\cal O}(\beta^2)$ relative to the $\beta=0$ case.

\end{itemize}

These observations offer the possibility to reduce the number of evolution equations by computing 
only the pure CDM or BM spectra and approximating the mixed ones. In the resulting equations for the pure spectra we make 
the following approximations:
\begin{itemize}
\item
In the r.h.s. of eqs.~(\ref{spectev1}) we approximate the mixed power spectra 
as the geometrical averages of the pure ones, i.e.
 \begin{align}\label{approxspectra}
 P_{13}(k,\eta)&\simeq\sqrt{P_{11}(k,\eta)P_{33}(k,\eta)},&
P_{14}(k,\eta)&\simeq\sqrt{P_{11}(k,\eta)P_{44}(k,\eta)},\nonumber\\
P_{23}(k,\eta)&\simeq\sqrt{P_{22}(k,\eta)P_{33}(k,\eta)},&
P_{24}(k,\eta)&\simeq\sqrt{P_{22}(k,\eta)P_{44}(k,\eta)}\;.
\end{align}
If the bias were $k$-independent, we would have $P_{13}=bP_{11}$, $P_{33}=b^2P_{11}$, and the first relation would
be exact. The same is true for the other relations, because of the $k$-independence of the linear growth functions.
The accuracy of the resulting mixed power spectra 
can be estimated through the $k$-dependence of the bias. As we have discussed above, this is an ${\cal O}(\beta^2)$ effect, 
which is smaller than 1\% for the values of $\beta$ that we use ($\beta \leq 0.1$). 
\item 
In the r.h.s. of eqs.~(\ref{spectev2}) the mixed components of the bispectra are approximated by the corresponding pure ones, e.g. 
\be \label{approxbi}
B_{113}\simeq B_{111}\,,
\ee
and so on. The direction of the approximation (whether a CDM index is turned into a BM one, or vice versa) is decided 
by a majority criterium: if a bispectrum has two CDM components and a BM one, 
it is approximated by a purely CDM bispectrum,  and vice versa. The accuracy of this approximation is determined by the 
magnitude of the bias. The deviation of the bias from 1 receives a correction around 1-2\% at low redshifts because of the
different initial conditions for CDM and BM, and a correction of ${\cal O}(\beta^2)$ because of the CDM-BM coupling $\beta\not=0$. 

We emphasize that this approximation affects the calculation of the power spectra only indirectly.
The approximated mixed bispectra appear in the r.h.s. of eq. (\ref{spectev2}) along with several pure ones. 
This induces an error in the pure bispectra, obtained through the integration of 
eq. (\ref{spectev2}),  which is significantly smaller than 1-2\%. 
The pure bispectra then affect the calculation of the power spectra by appearing in the r.h.s. of eq. (\ref{spectev1}).
We expect the residual effect of approximations such as (\ref{approxbi}) 
on the accuracy of the computed power spectra to be below 1\%.
\end{itemize}

Through the above approximations the system is reduced to a set of 22 coupled equations, which can be solved in a way analogous to 
that described in Appendix B of \cite{Max2}. As we have explained in detail, 
our approximations are expected to be valid at the sub-percent level for the 
power spectra within the BAO range. This can be verified a posteriori, by comparing with N-body simulations, 
as we will do in the following section. Of course, the most unambiguous test would be the comparison with an exact 
solution of the full 74 equations (\ref{spectev1}), (\ref{spectev2}). Unfortunately, it is very difficult to keep 
the numerical accuracy of the solution of such a huge system of integro-differential equations at the sub-percent 
level, in order to carry out this test.

\begin{figure}[t]
\begin{center}
\includegraphics[width=0.8\textwidth]{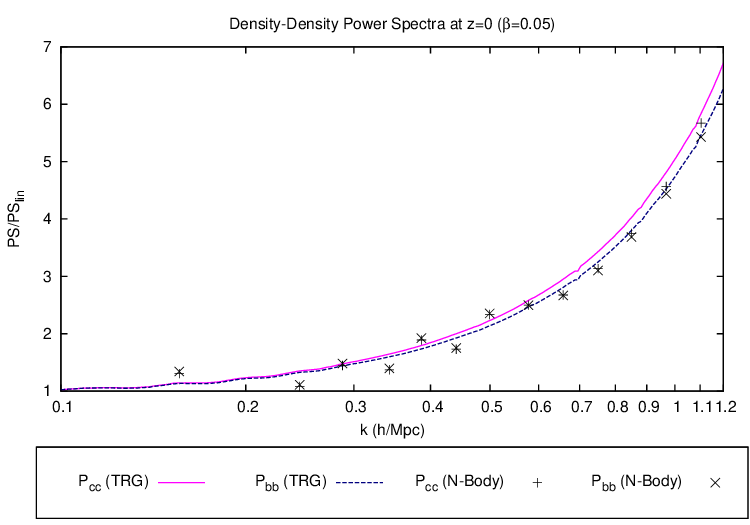}
\end{center}
\caption {Comparison of results from N-body simulations and our calculation ($\beta=0.05$). We
display the ratio of the non-linear and linear spectra for $z=0$.} 
\label{nbody05}
\end{figure}

\begin{figure}[t]
\begin{center}
\includegraphics[width=0.8\textwidth]{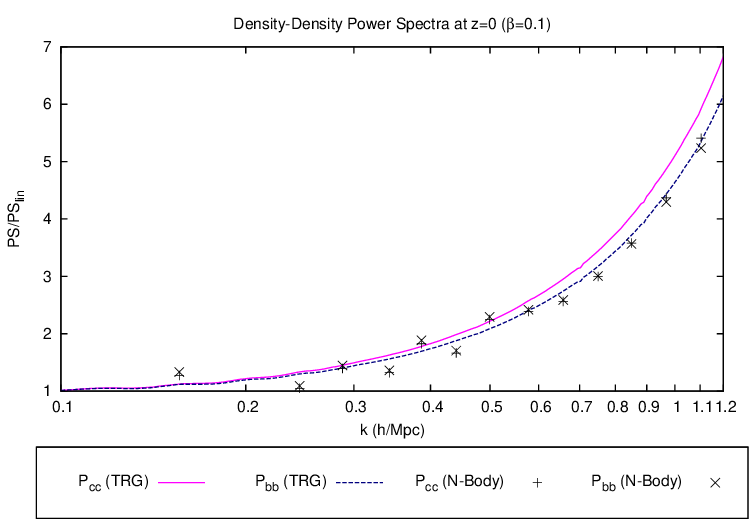}
\end{center}
\caption {Comparison of results from N-body simulations and our calculation ($\beta=0.1$). We
display the ratio of the non-linear and linear spectra for $z=0$.} 
\label{nbody1}
\end{figure}

\subsection{Results}

We have applied the formalism of the TRG to a quintessence model with non-zero coupling between CDM and the quintessence 
field $\phi$. We chose a particular model for which the matter spectrum has been calculated in \cite{baldi} 
through numerical N-body simulations.  The field has a potential $V(\phi) \sim \phi^{-\alpha}$, with $\alpha=0.143$. The 
present-day energy content of the Universe has $\Omega_{DE}=0.743$, $\Omega_{CDM}=0.213$, $\Omega_{BM}=0.044$.  
The Universe is assumed to have vanishing spatial curvature ($\Omega_k=0$), 
current expansion rate $H_0=71.9$ km s$^{-1}$ Mpc$^{-1}$. The mass variance is taken $\sigma_8=0.769$, as calculated 
from the linear spectrum.

The initial conditions for the integration of the evolution equations for the spectra have been set at a redshift $z=40$. At such 
early times the evolution is linear to a very good approximation. We employed the implementation of the
background and linear-perturbation equations in the Boltzmann code CMBEASY \cite{doran}, generalized for the interacting
case \cite{baldi}. We assumed that the primordial spectrum is scale invariant with spectral index $n=0.963$. We chose the 
initial value of the scalar field close to its tracker value in the uncoupled case, and adjusted the value of the dimensionful 
constant in its potential so as to obtain the present-day energy content listed above.

\begin{figure}[t]
\begin{center}
\includegraphics[width=0.8\textwidth]{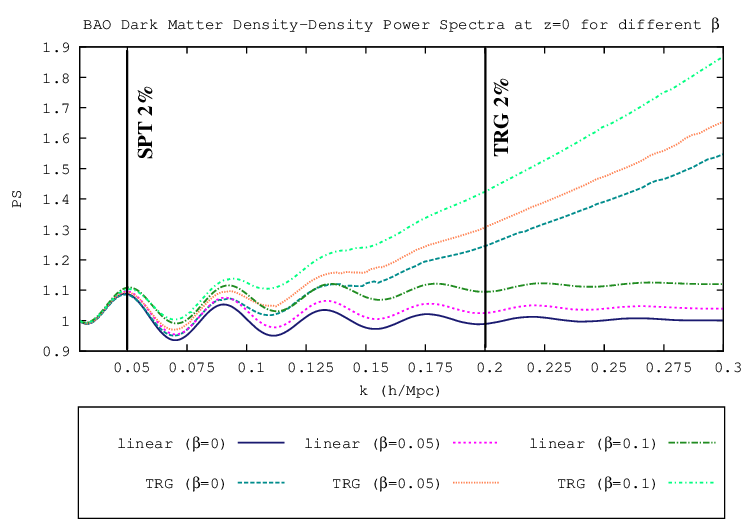}
\end{center}
\caption {Dark matter density-density spectra for various $\beta$ at $z=0$, normalized with respect to the smooth function of \cite{eisenstein}. The spectra have been multiplied by an additional $\beta$-dependent factor in order to be equal to 1 for $k\to 0$.
We indicate the maximum $k$ for which a certain level of accuracy is achieved by TRG and SPT.} 
\label{ddcc}
\end{figure}

In the ${\rm\Lambda CDM}$ case the momentum dependence of the power spectrum ensures that the 
momentum integrations in the evolution equations (\ref{spectev1}), (\ref{spectev2}) are both 
infrared (IR) and ultraviolet (UV) finite at any order in perturbation theory.
However, in order to perform numerical computations
these integrations  must be cut off both in the IR and the
UV. An appropriate IR limit 
eliminates contributions from very large length scales, of the order of the horizon distance. In our numerical study of the 
case of coupled quintessence we employ  
an IR cutoff $k_{IR}\simeq10^{-2}h/$Mpc. We have checked 
that an IR cutoff $k_{IR}\simeq10^{-3}h/$Mpc does not alter our results for the spectra. It tends, however, to increase the
noise for numerical integrations with larger time steps than the ones we employ. In principle, the UV cutoff must also be chosen with care.  Contributions with very large momenta correspond to length scales for which our method is not accurate. At scales below a few Mpc the process of virialization is crucial for the formation of galaxies and clusters of galaxies. The non-linear corrections that
we take into account through the TRG are not sufficient for the quantitative description of the physics at these scales.
For this reason we implement an UV cutoff $k_{UV}\simeq 2.3 \,h/$Mpc. We have checked 
that the variation of  $k_{UV}$  by a factor of two induces a variation of the spectra in the BAO range 
($0.03\,h$/Mpc $\lta k \lta 0.25\, h$/Mpc) at the sub-percent level. 
As a result, the evolution at small length scales
does not influence appreciably the evolution at the scales relevant for the BAO, which are the main focus of our 
calculation.

\begin{figure}[t]
\begin{center}
\includegraphics[width=0.8\textwidth]{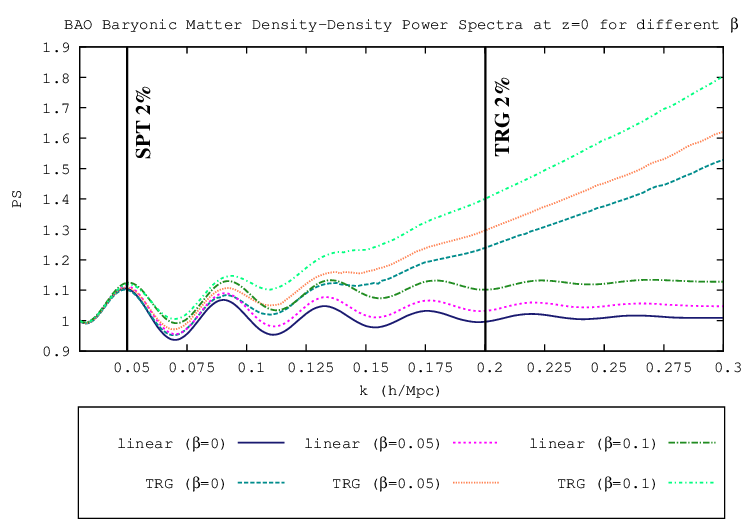}
\end{center}
\caption {Baryonic matter density-density spectra for various $\beta$ at $z=0$, normalized similarly to those in fig. \ref{ddcc}.} 
\label{ddbb}
\end{figure}

We can check the reliability of the TRG approach by comparing our results with those from N-body simulations.
In figs. \ref{nbody05} and \ref{nbody1} we display the ratio of non-linear to linear power spectra at $z=0$, for 
$\beta=0.05$ and 0.1 respectively. The lines correspond to our results for the CDM and BM density-density power
spectra. The points indicate the results of the N-body simulations presented in \cite{baldi} for the same quantities.
No error bars are given for these results that would permit an accurate assessment of the level of agreement
with our findings. On the other hand, we observe consistency of the two methods at
scales below $0.5\,h$/Mpc: 
The growth of the non-linear power spectrum relative to
the linear one at small length scales is captured well by the TRG. It is also obvious from the same figures that the two approaches 
give results that start to deviate for $k\gta 0.5\, h$/Mpc. This is expected, as the TRG cannot capture the 
processes of formation of bound structures that dominate at large momentum scales. Our main interest lies in
the region $0.03\,h$/Mpc $\lta k \lta 0.25\, h$/Mpc, in which the BAO are visible. We also point out that the
N-body simulations are rather noisy at such length scales because of finite volume effects, as they are 
optimized for scales smaller than the BAO range. 
For this reason, the two methods, N-body simulations and TRG, can be viewed as 
complementary: N-body simulations give a reliable description of the process of virialization at relatively small
length scales, while the TRG is the appropriate tool for the study of non-linear effects at the BAO range. 
The agreement of the two methods in the intermediate range of overlap is a confirmation their consistency.

\begin{figure}[t]
\begin{center}
\includegraphics[width=0.8\textwidth]{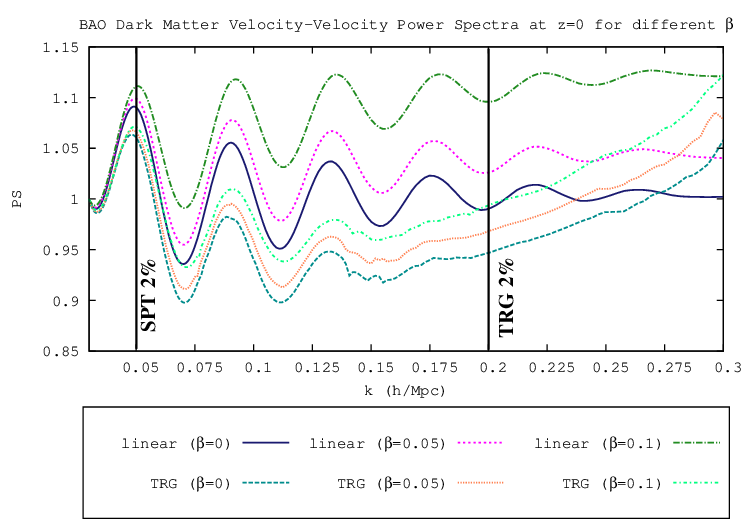}
\end{center}
\caption {Dark matter velocity-velocity spectra for various $\beta$ at $z=0$,normalized similarly to those in fig. \ref{ddcc}.} 
\label{vvcc}
\end{figure}

In figs. \ref{ddcc}, \ref{ddbb}, \ref{vvcc},  we display the linear and non-linear 
power spectra for models with $\beta=0$, 0.05 and 0.1.
We focus on the BAO range. In order to provide a clear depiction of the oscillatory behavior of the spectra we divide them
by the smooth function of ref. \cite{eisenstein} for $\Omega_{BM}=0$. 


In fig. \ref{ddcc} we display the density-density power spectra for CDM for various values of $\beta$. 
It is apparent that the inclusion of the non-linear effects causes the spectrum to increase at large momenta. 
For $k\lta 0.06\,h$/Mpc (a region that includes the first peak), the linear and non-linear spectra are indistinguishable. This indicates
that the non-linear corrections become negligible at large length scales. On the other hand, the non-linear spectrum becomes 
significantly larger than the linear one at smaller length scales. The non-linear effects are quantitatively 
important in the BAO range. In the vicinity of the second peak they provide corrections at the level of a few percent, while in the
vicinity of the third peak the corrections are about 10\%. These features of the non-linear corrections on the matter power spectrum in the $\beta=0$ case have been discussed in full detail in the past (for instance, in \cite{Max2}).

The presence of an extra CDM-DE coupling causes an additional enhancement of the spectrum. 
For example, for $k=0.3\, h$/Mpc the differences 
of the non-linear spectra for $\beta=0$, 0.05 and 0.1 have roughly doubled compared to the corresponding differences of
the linear spectra.
A qualitative change in the spectrum is that the third peak ceases to exist for $\beta \simeq 0.1$.
It is likely that the disappearance of higher-order peaks is common in models of coupled quintessence. 
However, it is not possible to investigate this feature in a model-independent way. The large variability of 
the evolution of the cosmological background 
and the perturbations in models of coupled quintessence makes it very difficult to identify generic features. 

In fig. \ref{ddbb} we display the density-density power spectra of BM. They are very similar to those of CDM. A close
inspection reveals that the enhancement of the spectrum with increasing $\beta$ is smaller than for the CDM case. This is
more apparent for $k=0.3\, h$/Mpc. The reason can be traced to the additional attractive force between CDM particles, 
mediated by the quintessence field. As the additional force is not felt by the baryons, 
the enhancement affects CDM and baryons differently,  i.e. a bias is produced. 
This is apparent already at the linear level, but it becomes an even stronger effect at the non-linear level. 
The baryons are not subject to this force and tend to collapse more slowly \cite{tocchini}. 
The bias is {\em scale dependent}, as we discuss in detail below.

In fig. \ref{vvcc}  we depict the velocity-velocity power spectra for CDM.  Similarly to the density-density
spectra, the higher-order peaks are washed away by the combined effect of the non-linear corrections and the CDM-DE interaction.
The non-linear density-density (dd), density-velocity (dv) and velocity-velocity (vv) 
power spectra display similar behavior.
The BM spectra follow closely the variation of the corresponding DM ones. There is also a hierarchy in the 
magnitude of the spectra for large $k$. The reason is that the velocity field is much smaller than the density 
perturbation at subhorizon scales. As a result, the non-linear effects that enhance the spectra at large $k$ are
more pronounced for the density-density spectra and smallest for the velocity-velocity ones. 

The bias in the BAO region is depicted in fig. \ref{bias1} for a redshifts $z=1.12$.
It is  defined as
$b(k)
= \left(P_{BM}(k)/P_{CDM}(k) \right)^{1/2}$.  
The small kinks in the curves give an indication of the numerical errors in our calculation.
We notice that, even at vanishing coupling ($\beta=0$), there is a bias 
as a result of the different initial conditions between BM and CDM after decoupling \cite{tocchini}. 
However, in the $\beta=0$ case the bias parameter tends to unity at lower redshifts, as is well known. 
When $\beta\neq 0$ the initial unbalance between baryons and CDM is never washed out, as a consequence 
of the extra scalar force acting on the latter and not on the former. In linear perturbation theory, 
this causes a {\it scale-independent} bias in recent epochs \cite{tocchini}. When non-linear corrections are taken 
into account, the situation changes dramatically. The CDM and BM components, having different initial conditions, 
evolve differently even in the $\beta=0$ case, with the BM fluid being always more ``linear" than the CDM one. 
Moreover, since the non-linear growth factors are scale-dependent, non-linear corrections cause a {\it scale-dependent} 
bias even at $\beta=0$ \cite{smith}. It is clear from fig. \ref{bias1} that both effects, 
 i.e. the bias and its scale dependence, are enhanced 
by a non-vanishing coupling. This feature could provide a distinctive signature for this kind of models.  
These results are consistent with the conclusions of refs. \cite{baldi,mainini3}, in which the emphasis is put on
the halo region of the spectrum.

\begin{figure}[t]
\begin{center}
\includegraphics[width=0.8\textwidth]{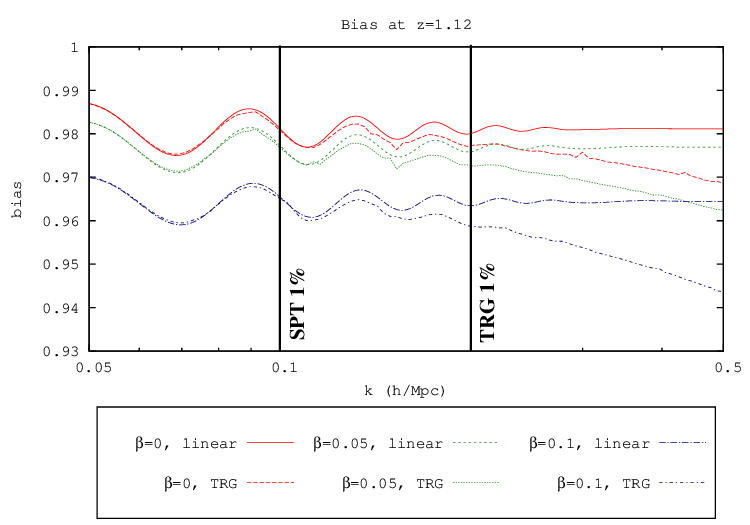}
\end{center}
\caption {The bias in the BAO region for $z=1.12$.} 
\label{bias1}
\end{figure}

\section{Conclusions}
\setcounter{equation}{0}

In this paper we have extended the TRG formalism introduced in ref.~\cite{Max2} in two respects.
Firstly, we described the matter sector of the theory keeping track of the CDM and BM components separately, 
instead of treating them as a single fluid. 
Secondly, we introduced a new scalar force that couples differently to CDM and BM. 
As was discussed recently in \cite{smith}, an accurate computation of the evolution of the different components is mandatory 
if one wants to achieve high precision modeling of structure formation. 
In order to test the accuracy of the TRG method for this more general class of cosmological scenaria, 
we analyzed the same cosmologies considered in \cite{baldi}, 
and compared our results with those of the N-body simulations presented there. 
The agreement is very good up to $k \simeq 0.5 \,\mathrm{h/Mpc}$ at $z=0$, where the 
non-linear power spectrum is roughly twice the linear one. These results confirm the reliability of the TRG as 
a computational tool that can fill the gap between linear and non-linear scales.

Even in the absence of an extra force, 
or in the case that the extra force couples universally to all matter species, 
the evolution of BM and CDM differs as a consequence of different initial conditions after decoupling. 
In linear theory, the initial unbalance between BM and CDM fluctuations is almost completely washed out by the present epoch. However, when non-linearities are taken into account the bias persists. We have seen that the effect is at the percent 
level in the BAO range at low redshift in the uncoupled case, 
but it may grow up to the $2.5\%$ level when a non-zero coupling is turned on with a value compatible with 
present bounds (obtained within the linear approximation \cite{bean,lavacca}). As a result, these models will receive 
significant constraints from future galaxy surveys, which aim to measure the power spectrum within the BAO range with
an accuracy at the percent level. 
A thorough investigation of this model-dependent issue goes beyond the purpose of this paper and is postponed for future work.

The most interesting outcome of our analysis from the point of view of observations 
concerns the form of the bias between CDM and BM spectra. In linear theory, the
bias is {\it scale independent} in recent epochs \cite{tocchini}. The non-linear corrections remove this feature, and make the
bias {\it scale dependent}, consistently with the results of \cite{baldi}. We confirm and extend these results
for the BAO region ($0.03\,h$/Mpc $\lta k \lta 0.25\, h$/Mpc), where we expect the TRG method to be reliable. 
The effect becomes more pronounced with increasing coupling 
between CDM and DE (assuming that the respective coupling for BM vanishes). 
The form of the bias provides an additional observational handle for the differentiation 
between various models. For example, in models with massive neutrinos the growth of the spectrum induced by the non-linearities
at small length scales is compensated by the free-streaming of neutrinos \cite{lesgourgues}. 
On the other hand, the bias is expected to remain
scale-dependent at the non-linear level if there is a substantial CDM-DE coupling.

\section*{Acknowledgments}
F.~S. would like to thank C. Nardini and D. Seminara for useful discussions. We are grateful to 
M. Baldi for providing the N-Body data of \cite{baldi}
for comparison with our results.
M.~P. and N.~T. are supported in part by the EU Marie Curie Network ``UniverseNet'' 
(MRTN--CT--2006--035863).  N.~T is also supported in part by the ITN network
``UNILHC'' (PITN-GA-2009-237920).
V.~P. acknowledges support from the Alexander von  
Humboldt Foundation.
G.~R. is supported by the German Aeronautics Center and Space Agency (DLR),
under program 50-OP-0901, funded by the Federal Ministry of Economics and 
Technology.


\section{Appendix A: Non-linear evolution of perturbations}
\label{formalism}
\setcounter{equation}{0}

We consider first the possibility that the energy density of the Universe 
is dominated by two components:  
a) a species of weakly interacting, massive particles, which we identify with
dark matter (CDM), and  
b) a slowly varying, classical scalar field $\phi$, whose contribution is 
characterized as dark energy (DE).
There is a direct coupling between the particles and the 
scalar field:
The mass $m$ of the particles depends on
the value of $\phi$. 
For classical particles, the action of the system can be written as 
\begin{equation}
{\cal S}=\int d^4x \sqrt{-g}
\left(\frac{1}{16\pi G} R+\frac{1}{2}
\frac{\partial \phi}{\partial x^\mu}
\frac{\partial \phi}{\partial x^\nu}
g^{\mu\nu}-U(\phi) \right)
-\sum_i \int m(\phi(x_i))ds_i,
\label{action} \end{equation}
with $ds_i=\sqrt{
g_{\mu\nu}(x_i)dx^\mu_idx^\nu_i}$ and
the second integral taken over particle trajectories. 
Variation of the action with respect to $\phi$ results in the
equation of motion 
\begin{eqnarray}
\frac{1}{\sqrt{-g}}\frac{\partial}{\partial x^\mu}
\left(\sqrt{-g}\,\,g^{\mu\nu}\frac{\partial \phi}{\partial x^\nu}
\right)&=&
-\frac{dU}{d\phi}-
\frac{1}{\sqrt{-g}}\sum_i\int ds_i \,\, 
\frac{dm(\phi(x_i))}{d\phi} \, \delta^{(4)}(x-x_i)
\nonumber \\
&=&
-\frac{dU}{d\phi}+\frac{\beta(\phi)}{M}\,\, \left( T_{CDM} \right)^\mu_{~\mu},
\label{eomphi} \end{eqnarray}
with the energy-momentum tensor associated with the gas of particles
given by  
\be
\left( T_{CDM} \right)^{\mu\nu}=\frac{1}{\sqrt{-g}} 
\sum_i \int ds_i \,\, m(\phi(x_i))\,\, 
\frac{dx_i^\mu}{ds_i}\frac{dx_i^\nu}{ds_i}
\delta^{(4)}(x-x_i).
\label{emtg} \ee
Here $M$ is the reduced Planck mass $M=(8\pi G)^{-1/2}$ and 
we have defined $\beta(\phi)/M=-{d\ln m(\phi)}/{d\phi}$. 
In the following we shall use eq. (\ref{eomphi}), but we shall approximate the 
energy-momentum tensor as that of an ideal pressureless fluid.
The same equation is obtained for scalar-tensor theories of gravity in 
the Einstein frame. If more than one massive species are present in such theories,
their coupling to the scalar field is universal. 
In order to be as general as possible when we discuss multiple massive species in the following appendix, we assume that each species 
has a different coupling $\beta_i$, induced through the $\phi$-dependence of 
its mass.

For the metric, we consider an ansatz of the form (\ref{metric}).
We assume that the Newtonian potential $\Phi$ is weak, $\Phi \ll 1$, and  
that the field $\phi$ can be decomposed as in eq. (\ref{decomphi}),
with $\dphi/\bar{\phi} \ll 1$. In general,
$\bar{\phi}={\cal O}(M)$.
The magnitude of the fluctuations of $\phi$ is 
expected to be similar to that of the gravitational field $\Phi$.
The reason is that the source for both is the dark matter density, to 
which they couple with comparable strength. 
Finally, the density can be decomposed as in eq. (\ref{decomrho}).
Our aim is to take into account the effect of the local velocity field
$\dvvec$,
when this becomes significant because of large field gradients. For 
subhorizon perturbations with momenta $k\gg \Hc=\dot{a}/a$, 
the linear analysis predicts 
$\dvvec \sim (k/\Hc) \Phi \sim (\Hc/k)(\drho/\bar{\rho})$. We assume
that these relations are approximately valid even at the non-linear level, within
the range of validity of our scheme.
Our assumptions can be summarized in the  
hierarchy of
scales: $\Phi, \dphi/\bar{\phi} \ll |\dvvec | \ll \drho/\bar{\rho}  \lesssim 1$. 
At the linear level, 
we have $\dvvec^2 \sim \Phi (\drho/\bar{\rho})$. We assume that such a relation
holds at the non-linear level as well.
As we are dealing
with subhorizon perturbations, we expect that the spatial derivatives of $\Phi, \dphi$
dominate over their time derivatives. 
Following linear theory,
we make the more specific assumption 
that a spatial derivative acting on $\Phi$, $\dphi$ or 
$\dvvec$ increases the position of that quantity in the hierarchy by one 
level. In this sense $\vec{\nabla} \Phi$ is comparable to $\dvvec$, while
$\nabla^2 \Phi$ is comparable to $\bar{\rho}$.

We approximate the energy-momentum tensor of dark matter as 
$ \left( T_{CDM} \right)^{\mu\nu}= \rho\, V^\mu V^\nu$. 
We define the peculiar velocity through $V^i=\dv^i/a$.
Keeping the leading terms in our expansion gives
\begin{eqnarray}
\left( T_{CDM} \right)^0_{~0}&=&\rho(1+\dvvec^2)
\nonumber \\
\left( T_{CDM} \right)^0_{~i}&=&-\rho\,\dv^i
\nonumber \\
\left( T_{CDM} \right)^i_{~j}&=&-\rho\,\dv^i\,\dv_j,
\label{tdm} \end{eqnarray}
with $\dv_j=\dv^j$.
We emphasize at this point that our assumption for the form of 
$\left( T_{CDM} \right)^{ij}$ is consistent with the presence of only
one gravitational potential $\Phi$ in our ansatz (\ref{metric}) for the metric, 
within the order that this potential will be determined through the Einstein
equations.

The energy-momentum tensor of the scalar field has the 
leading part
\begin{eqnarray}
\left( T_{S} \right)^0_{~0}&=&\frac{1}{2a^2}\dot{\bar{\phi}}^2+U(\bar{\phi})
\nonumber \\
\left( T_{S} \right)^0_{~i}&=&0
\nonumber \\
\left( T_{S} \right)^i_{~j}&=&
\left[-\frac{1}{2a^2}\dot{\bar{\phi}}^2+U(\bar{\phi}) \right]\delta^i_{~j},
\label{sc0} \end{eqnarray}
with a dot denoting a derivative with respect to $\tau$. It also includes 
a perturbation 
\begin{eqnarray}
\left(\delta T_{S} \right)^0_{~0}&=&
\frac{1}{a^2}\left[ -\dot{\bar{\phi}}^2 \Phi+\dot{\bar{\phi}}\delta\phidot+
\frac{1}{2}( \vec{\nabla} \dphi )^2  +a^2\frac{dU(\bar{\phi})}{d\phi}\delta\phi
\right]
\nonumber \\
\left( \delta T_{S} \right)^0_{~i}&=&
\frac{1}{a^2}\dot{\bar{\phi}} \delta\phi_{,i}
\nonumber \\
\left(\delta T_{S} \right)^i_{~j}&=&
\frac{1}{a^2}\left[ \dot{\bar{\phi}}^2 \Phi-\dot{\bar{\phi}}\delta\phidot
+\frac{1}{2}( \vec{\nabla} \dphi )^2  +
a^2\frac{dU(\bar{\phi})}{d\phi}\delta\phi
\right]\delta^i_{~j}-\frac{\partial^{i}(\delta\phi)\partial_{j}(\delta\phi)}{a^{2}}.
\label{scpert} \end{eqnarray}
Notice that we have included a term $( \vec{\nabla} \dphi )^2/2  $, as it
is comparable to $\dvvec^2$ or $\Phi$ within our assumed hierarchy. 
The same is not true for $(\delta\dot{\phi})^2/2$, which is subleading.

The equation of motion of the scalar field (\ref{eomphi}) can be split into an
equation for the homogeneous part:
\be
\ddot{\bar{\phi}}+2\frac{\dot{a}}{a}\dot{\bar{\phi}}+a^2\frac{dU(\bar{\phi})}{d\phi}=
+\frac{\beta(\bar{\phi})}{M}a^2\bar{\rho},
\label{homphi} \ee
and one for the perturbation:
\be
\delta\ddot{\phi}+2\frac{\dot{a}}{a}\delta\dot{\phi}-\nabla^2\delta
\phi+a^2\frac{d^2U(\bar{\phi})}{d\phi^2}\dphi-4\dot{\bar{\phi}}\dot{\Phi}
+2 a^2\frac{dU(\bar{\phi})}{d\phi}\Phi
=\frac{\beta(\bar{\phi})}{M}a^2\drho
+\frac{1}{M}\frac{d\beta(\bar{\phi})}{d\phi}a^2(\bar{\rho}+\drho)\ \dphi.
\label{pertphi} \ee
The terms with
time derivatives are subdominant according to our assumptions.
Neglecting them results in 
\be
-\nabla^2\delta\phi
+\left[a^2\frac{d^2U(\bar{\phi})}{d\phi^2}
-\frac{1}{M}\frac{d\beta(\bar{\phi})}{d\phi}(\bar{\rho}+\drho)\right] \dphi
+2 a^2\frac{dU(\bar{\phi})}{d\phi}\Phi
=\frac{\beta(\bar{\phi})}{M}a^2\drho.
\label{pertphii} \ee
For the scalar field $\bar{\phi}$ 
to evolve at cosmological times, it must have a mass term
$d^2U(\bar{\phi})/d{\phi}^2 = {\cal O} (\Hc^2)$. We also have
$U(\bar{\phi}),\, \bar{\rho} =  {\cal O} (\Hc^2 M^2)$. It is natural to expect 
$dU(\bar{\phi})/d{\phi} = {\cal O} (\Hc^2 M)$. For subhorizon perturbations
with momenta $k\gg \Hc$, neglecting the subleading terms 
results in 
a very simple Poisson equation for the 
field $\dphi$:
\be
\nabla^2\dphi=-\frac{\beta(\bar{\phi})}{M} a^2\drho.
\label{phipois} \ee

The equation of motion for the gravitational potential $\Phi$ can be obtained from
the first Einstein
equation. The leading terms give 
\be
\Hc^2=\left( \frac{\dot{a}}{a} \right)^2=
\frac{1}{3 M^2}\left[ a^2 \bar{\rho}+\frac{1}{2}\dot{\bar{\phi}}^2+a^2 U(\bar{\phi}) 
\right],
\label{hubble} \ee
while the equation for the perturbation is
\be
\nabla^2 \Phi -3 \Hc\dot{\Phi}-3\Hc^2\Phi=\frac{1}{2M^2}
\left[a^2\drho+a^2\bar{\rho} \,\dvvec^2
-\dot{\bar{\phi}}^2\Phi+\dot{\bar{\phi}}\delta\dot{\phi}+a^2\frac{dU(\bar{\phi})}{d\phi}\dphi
\right].
\label{gravpot}
\ee
Our assumptions about the hierarchy of the various scales and 
the dominance of the spatial derivatives
lead to the Poisson equation for
the gravitational field $\Phi$:
\be
\nabla^2\Phi=\frac{1}{2M^2} a^2 \drho.
\label{gravpois} \ee


We now turn to equations derived from the conservation of the total
energy-momentum tensor $T^{\mu\nu}_{~~;\nu}=0$. For $\mu=0$, the
leading terms give
\be
\dot{\bar{\rho}}+3\Hc \bar{\rho}
=-\frac{\dot{\bar{\phi}}}{a^2}
\left( \ddot{\bar{\phi}}+2\Hc \dot{\bar{\phi}}+a^2\frac{dU(\bar{\phi})}{d\phi}
\right)=
-\frac{\beta(\bar{\phi})}{M}\dot{\bar{\phi}} \bar{\rho},
\label{cons2} \ee
where we have employed eq. (\ref{homphi}).
The equation for the perturbations is more complicated. It can be simplified 
considerably through
our assumptions about the hierarchy of the various fields. We obtain
\be
\delta\dot{\rho}+3\Hc\drho+\vec{\nabla}\left[(\bar{\rho}+\drho)\dvvec
\right]= \frac{\dot{\bar{\phi}}}{a^2} \nabla^2 \dphi=
-\frac{\beta(\bar{\phi})}{M}\dot{\bar{\phi}}\,\drho,
\label{cons3} \ee
where we have employed eq. (\ref{phipois}).

For $\mu=i=1,2,3$ we obtain the generalization of the Euler equation for this
system. After eliminating higher-order terms we find
\begin{eqnarray}
(\bar{\rho}+\drho\ )\left[
\delta \dot{\vec{v}}+\Hc \dvvec+\left(\dvvec\, \vec{\nabla}\right)\dvvec
\right]
&+& \Bigl[ \dot{\bar{\rho}}+3\Hc\bar{\rho} 
+ \delta\dot{\rho}+3\Hc\drho+\vec{\nabla}\left[(\bar{\rho}+\drho)\dvvec
\right] \Bigr] \dvvec 
\nonumber \\
=-\left(\bar{\rho}+\drho\right)\vec{\nabla} \Phi
&+&\frac{1}{a^2}
\left( \ddot{\bar{\phi}}+2\Hc\dot{\bar{\phi}}-\nabla^2\dphi+a^2\frac{dU(\bar{\phi})}{d\phi}
\right)\vec{\nabla} \dphi.
\label{cons4} \end{eqnarray}
Employing eqs. (\ref{homphi}), (\ref{phipois}), (\ref{cons2}), (\ref{cons3}) we find
\be
\delta \dot{\vec{v}}
+\left(\Hc- \frac{\beta(\bar{\phi})}{M}
\dot{\bar{\phi}}
\right) \dvvec
+\left(\dvvec\, \vec{\nabla}\right)\dvvec
=-\vec{\nabla} \Phi
+\frac{\beta(\bar{\phi})}{M}\vec{\nabla} \dphi.
\label{cons5} \ee

For constant $\beta$,
we can form a combination of eqs. (\ref{phipois}), (\ref{gravpois}) of the form
\be
\nabla^2\Phit=\frac{1}{2\Mt} a^2 \drho,
\label{dm3} \ee
where 
$\Phit=\Phi-\beta\dphi/M$ and $\Mt^2=M^2/(1+2\beta^2)$.
Eq. (\ref{cons5}) becomes
\be
\delta \dot{\vec{v}}
+\left(\Hc - \frac{\beta}{M}\dot{\bar{\phi}}
\right) \dvvec
+\left(\dvvec\,\cdotp \vec{\nabla}\right)\dvvec
=-\vec{\nabla} \Phit.
\label{dm66} \ee
We can see that the Newtonian potential
for the perturbations involves a stronger Newton's constant 
$\Gt=(8\pi\Mt^2)^{-1}=(1+2\beta^2)G$. 
There is also a correction $\sim \dot{\bar{\phi}} \dvvec$ in the l.h.s. of the
Euler equation, because the particles do not follow geodesic motion with
respect to the background metric.\\

\section{Appendix B: Several species of non-relativistic matter}
\setcounter{equation}{0}
In this appendix we generalize
the formalism to the case that there are more than one 
particle species contributing significantly to the energy density. Each
of them couples to the scalar field with a different coupling $\beta_i$. 
One example of particular interest includes CDM with $\beta_{CDM}\not=0$ and baryonic
matter (BM), for which we assume that $\beta_{BM}=0$ in order to be consistent with
observational constraints. We also normalize all dimensionful quantities in terms
of the Planck mass. This is equivalent to setting $M=1$ in the expressions of
appendix A.

The background equations are
\begin{eqnarray}
\Hc^2&=&
\frac{1}{3}\left[ a^2 \sum_i\bar{\rho_{i}}+\frac{1}{2}\dot{\bar{\phi}}^2
+a^2 U(\bar{\phi}) 
\right]
\label{b1} \\
\dot{\bar{\rho}}_{i}+3\Hc \bar{\rho_{i}}&=&
-\dot{\bar{\phi}}{\beta_i} \bar{\rho_{i}}.
\label{b2} \\
\ddot{\bar{\phi}}+2\Hc\dot{\bar{\phi}}+a^2\frac{dU(\bar{\phi})}{d\phi}
&=&
a^2\sum_i{\beta_i}\bar{\rho_{i}},
\label{b3} \end{eqnarray}
with the index $i$ counting the various species.

For the perturbations, eq. (\ref{phipois}) becomes
\be
\nabla^2\dphi=-a^2 \sum_i {\beta_i} \drho_i,
\label{dm1} \ee
and eq. (\ref{gravpois}) 
\be
\nabla^2\Phi=\frac{1}{2} a^2 \sum_i \drho_i.
\label{dm2} \ee
Eq. (\ref{cons3}) generalizes to 
\be
\delta\dot{\rho}_i+3\Hc\drho_i+\vec{\nabla}\left[(\bar{\rho_{i}}+\drho_i)\dvvec_i
\right]=-{\beta_i}\dot{\bar{\phi}} \drho_i.
\label{dm4} \ee
Finally, eq. (\ref{cons5}) becomes
\be
\delta \dot{\vec{v}}_i
+\left(\Hc - {\beta_i}\dot{\bar{\phi}}
\right) \dvvec_i
+\left(\dvvec_i\,\cdotp \vec{\nabla}\right)\dvvec_i
=-\vec{\nabla} \Phi+{\beta_i}\vec{\nabla} \dphi.
\label{dm6} \ee

We can write the above equations in a more useful form
by defining the density contrasts 
$\delta_{i}\equiv{\delta\rho_{i}}/{\bar{\rho_{i}}}\lesssim 1$
and 
$\theta_{i}(\textbf{k}, \tau)\equiv\vec{\nabla}\cdot\vec{\delta v_{i}}(\textbf{k}, \tau)$.
For the density contrasts we obtain
\begin{equation}\label{stpa}
\dot{\delta}_{i}+\vec{\nabla}\big[(1+\delta_{i})\vec{\delta v}_{i}\big]=0.
\end{equation}
For the Fourier transformed quantities, 
eq. (\ref{stpa}) gives eq. (\ref{delta}), while 
eq. (\ref{b}) gives eq. (\ref{theta}).

We replace time by the variable $\eta=\ln a(\tau)$.  
For $n$ species of non-relativistic matter we
define the field $\varphi(\textbf{k}, \eta)$ as a vector with $2n$ components:
\begin{displaymath}\label{field}
 \varphi(\textbf{k}, \eta)=\left(
\begin{array}{c}
\varphi_{1}(\textbf{k}, \eta)\\ \\ \varphi_{2}(\textbf{k}, \eta)\\ \\ \vdots \\ \\ 
\varphi_{2n-1}(\textbf{k}, \eta)\\ \\ \varphi_{2n}(\textbf{k}, \eta)\end{array}
\right)
=e^{-\eta}\left(
\begin{array}{c}
\delta_{1}(\textbf{k}, \eta)\\ \\-\dfrac{\theta_{1}(\textbf{k}, \eta)}{\mathcal{H}}\\ \\ 
\vdots \\ \\
\delta_{n}(\textbf{k}, \eta)\\ \\-\dfrac{\theta_{n}(\textbf{k}, \eta)}{\mathcal{H}}
\end{array}
\right).
\end{displaymath}
This allows us to bring 
eqs. (\ref{delta}), (\ref{theta}) 
in the usual form \cite{CrSc1,Max1,Max2}: 
\begin{equation}\label{argh}
\partial_{\eta}\varphi_{a}(\textbf{k}, \eta)+\Omega_{ab}\varphi_{b}(\textbf{k}, \eta)
=e^{\eta}\gamma_{abc}(\textbf{k},-\textbf{k}_{1}, -\textbf{k}_{2})
\varphi_{a}(\textbf{k}_{1}, \eta)\varphi_{b}(\textbf{k}_{2}, \eta).
\end{equation}
The indices $a,b,c$ take values $1,\ldots, 2n$. 
Repeated momenta are integrated over while repeated indices are summed over. 
The functions $\gamma$, that determine effective vertices, 
are analogous to those employed in \cite{Max1,Max2}.
We find that the
non-zero functions are
\begin{equation}
\begin{split}
\gamma_{2i-1,2i,2i-1}(\textbf{k},\textbf{k}_{1}, \textbf{k}_{2})&
=\dfrac{\alpht(\textbf{k}_{1}, \textbf{k}_{2})}{2}\delta_{D}(\textbf{k}
+\textbf{k}_{1}+\textbf{k}_{2})=\gamma_{2i-1,2i-1,2i}(\textbf{k}, \textbf{k}_{2}, 
\textbf{k}_{1})\\
\gamma_{2i,2i,2i}(\textbf{k},\textbf{k}_{1}, \textbf{k}_{2})&=
\bett(\textbf{k}_{1}, \textbf{k}_{2})\ \delta_{D}(\textbf{k}+\textbf{k}_{1}
+\textbf{k}_{2}).
\end{split}
\label{vertices}\end{equation}

The $\Omega$-matrix of eq. (\ref{argh}) is a $2n\times2n$ matrix. 
Let us define the $2\times2$ matrices $\omega_{i}(\eta)$ and 
$\omega_{i,j}(\eta)$, with $i,j=1, \ldots, n$ and $i\neq j$, as 
\begin{equation}
 \omega_{i}(\eta)=\left(
\begin{array}{cc}
 1 & -1 \\ \\
-\dfrac{3}{2}\Omega_{i}(2\beta_{i}^{2}+1) & 2-\beta_{i}\bar{\phi}'
+\dfrac{\mathcal{H}'}{\mathcal{H}}
\end{array}
\right)
\label{omegai} \end{equation}
and 
\begin{equation}
 \omega_{i,j}(\eta)=\left(
\begin{array}{cc}
 0 & 0 \\ \\
-\dfrac{3}{2}\Omega_{j}(2\beta_i\beta_{j}+1) & 0
\end{array}
\right),
\label{omegaij} \end{equation}
where a prime denotes a derivative with respect to $\eta$.
Then, the $\Omega$-matrix can be written as
\begin{equation}
 \Omega(\eta)=\left(
\begin{array}{cccc}
 \omega_{1} & \omega_{1,2} & \ldots & \omega_{1,n }\\ \\
\omega_{2,1} & \omega_{2} & \ldots & \omega_{2,n }\\ \\
\vdots &  & \ddots & \\ \\
\omega_{n,1} & \omega_{n,2} & \ldots & \omega_{n}
\end{array}
\right).
\label{omegamatrix} \end{equation}
Notice that the only way in which different species of matter influence each other is 
through the matrices $\omega_{i,j}$, while the vertices 
do not mix contributions from different species.



\end{document}